\documentclass[aps,pre,twocolumn,showpacs,superscriptaddress,groupedaddress,floatfix,nofootinbib]{revtex4-2}
\usepackage{amsmath}
\usepackage{amssymb}
\usepackage{amsfonts}
\usepackage{graphicx}
\usepackage{dcolumn}
\usepackage{sidecap}
\usepackage{parskip}
\usepackage{grffile}
\usepackage{color}
\usepackage[breaklinks]{hyperref}
\usepackage{hhline}
\usepackage{mathtools}
\usepackage{graphics}
\usepackage{multirow}
\usepackage{verbatim}
\usepackage{longtable}
\usepackage{rotating}
\usepackage{setspace}
\usepackage{epsfig}
\usepackage{subfigure}
\usepackage{epstopdf}
\usepackage[normalem]{ulem}
\makeatletter
\def\@eqnnum{{\normalsize \normalcolor (\theequation)}}
 \makeatother

\graphicspath{{./}{ER/main/}}

\begin{document}
%------------------------------

\title{First-order route to antiphase clustering in adaptive simplicial complexes}
\author{Ajay Deep Kachhvah}
\email{iajaydeep@gmail.com}
\author{Sarika Jalan}
\email{sarikajalan9@gmail.com}
\affiliation{Complex Systems Lab, Department of Physics, Indian Institute of Technology Indore - Simrol, Indore - 453552, India}

%\date{\today}

\begin{abstract}
This Letter investigates the transition to synchronization of oscillator ensembles encoded by simplicial complexes in which pairwise and higher-order coupling weights alter with time through a rate-based adaptive mechanism inspired by the Hebbian learning rule. These simultaneously evolving disparate adaptive coupling weights lead to a phenomenon in which the in-phase synchronization is completely obliterated; instead, the anti-phase synchronization is originated. In addition, the onsets of antiphase synchronization and desynchronization are manageable through both dyadic and triadic learning rates. The theoretical validation of these numerical assessments is delineated thoroughly by employing Ott-Antonsen dimensionality reduction. The framework and results of the Letter could help in the understanding of the underlying synchronization behavior of a range of real-world systems, such as brain functions and social systems where interactions evolve with time.
\end{abstract}

%\pacs{89.75.Hc, 02.10.Yn, 5.40.-a}

\maketitle

\paragraph*{Introduction.} The inclusion of higher-order interactions has not drawn much attention for so long while envisaging the underlying dynamics influencing distinct processes taking place on a variety of complex systems ranging from physical to biological systems. Nevertheless, many complex systems, such as brain networks~\citep{Petri2014, Benson2018} and social interaction networks~\cite{Iacopini2019, Matamalas2020}, have the underlying structure of higher-order connections, which can be exemplified by simplicial complexes~\cite{Salnikov2018, Whitehead1939}. These higher-order interactions can be encoded by simplicial complexes, which are sets of $n$-simplexes, filled cliques of $n+1$ nodes, viz., vertices (0-simplex), lines (1-simplex), triangles (2-simplex), tetrahedrons (3-simplex), etc. An $n$-simplicial complex comprises the $n$-simplexes and the downward closure $(n{-}1)$-simplexes. Recently, the call for simplicial complexes in encoding higher-order interaction in complex systems has led to a sudden increase in untangling the reciprocation between network geometry and dynamical processes~\cite{Tanaka2011,Skardal2019,Xu2020,Millan2020,Skardal2020,Lucas2020,Battiston2020,Ghorbanchian2021,Xu2021,Chutani2021,Sun2021}. One novel phenomenon that naturally results from simplicial complex encoded higher-order interactions is the abrupt transition to synchronization and desynchronization~\cite{Skardal2019,Skardal2020,Kuehn2021}. Simplicial complexes are a suitable candidate for capturing the underlying geometry of complex systems; for instance, they have been used to encode the topological map of the environment's geometrical features captured by the hippocampus~\cite{Dabaghian2012}. 

The role of adaptation is instrumental in the growth and proper functioning of many physical and biological systems. For instance, it is a widespread perception in neuroscience that synaptic plasticity among the firing neurons forms the basis for the learning process and memory storage in the brain~\cite{Shimizu2000,Abbott2000}. It was Hebb~\cite{Hebb1949} who first put forth the concept that the simultaneous firing of the interacting neurons strengthens the synaptic connectivity between them~\cite{Abbott2000,Markram1997,Zhang1998}.\ Spike-timing-dependent plasticity between the firing neurons is one popular approach to understanding the impact of synaptic plasticity on various processes transpiring in the brain~\cite{Zhigulin2003, Knoblauch2012, Zirkle2020}. Nevertheless, the correlation between presynaptic and postsynaptic spike-timings of the interacting neurons is also encoded in phases of the oscillators to realize a neural network with synaptic plasticity.\ Such rate-based models of synaptic plasticity between the interacting neurons have divulged riveting structures and processes, for instance, cluster synchronization~\cite{Niyogi2009,Aoki2009,Gutierrez2011,Pitsik2018,Khanra2022,Shepelev2021} and abrupt synchronization and desynchronization~\cite{AGaytan2018,Kachhvah2020,Berner2021,Frolov2021,Majhi2022} in monolayer and multilayer networks. In cluster synchronization, a network is segregated into distinct clusters of nodes in which the nodes of the same cluster are mutually synchronized; still, the distinct clusters are not mutually synchronized. A diverse range of real-world systems exists, such as the cortical brain network~\cite{Lodi2020}, the power grid network~\cite{Pecora2014}, consensus dynamics~\cite{Schaub2016}, and schools of fish and swarms of birds~\cite{Hemelrijk2012}, having cluster synchronization as a key mechanism of their evolution or functioning.

\begin{figure}[t!]
	\centering
	\begin{tabular}{cccc}
	\includegraphics[height=3.5cm,width=8.5cm]{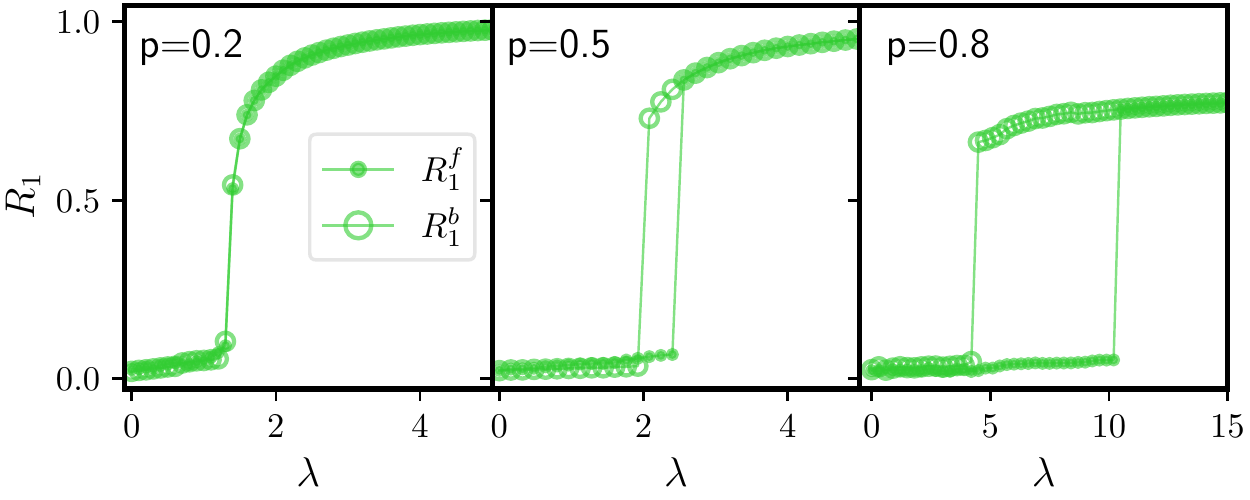}\\
	\includegraphics[height=3.5cm,width=8.5cm]{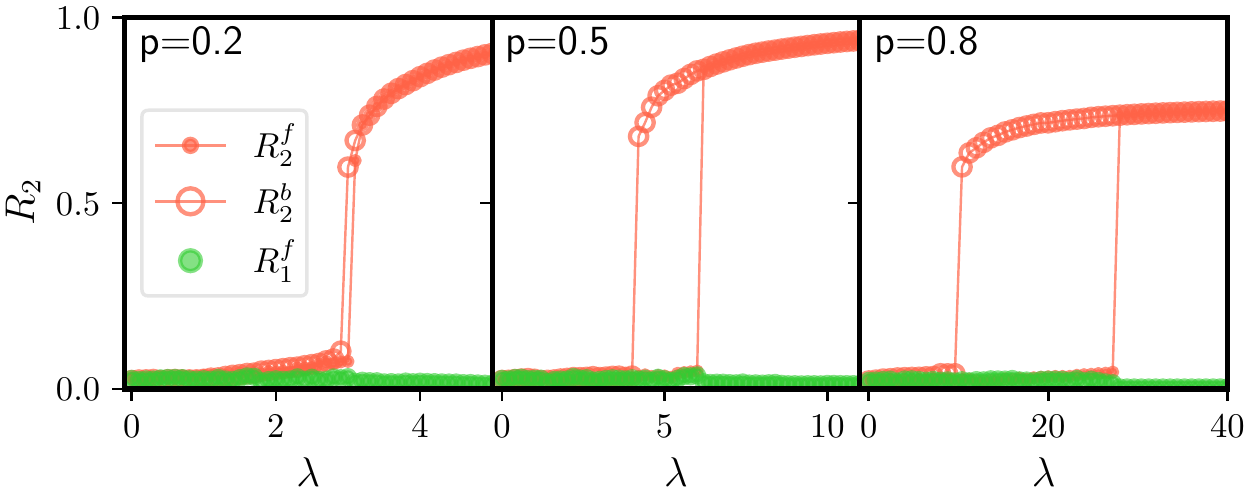}\\
	\end{tabular}{}
	\vspace{-0.5cm}
	\caption{(Color online) {Antiphase synchronization stems from adaptive couplings:} $R_1{-}\lambda$ (top row) and $R_2{-}\lambda$ (bottom row) profiles corresponding to the static couplings ($A_{ij}{=}1$ and $B_{ijk}{=}1$) and the adaptive couplings [Eq.~(\ref{weight_adapt})] with learning rates $\alpha{=}\beta{=}1$, respectively, in the random 2-simplicial complex.}
	\label{figure1}
\end{figure}
This Letter focuses on the impact of simultaneous adaptation of different simplex couplings on the transition to synchronization and desynchronization in simplicial complexes. Here the adaptation of 1-simplex (dyadic) and 2-simplex (triadic) couplings in a simplicial complex is inspired by the Hebbian learning rule, i.e., the dyadic and triadic weights are strengthened (weakened) if the dyad and triad of the oscillators establishing the respective connectivities are in phase (out of phase), respectively. Such concurrent adaptation in simplicial complexes leads to a fascinating phenomenon of abrupt antiphase synchronization while the in phase synchronization is completely inhibited.\ Moreover, the proposed model allows us to determine the onset of synchronization through the learning parameters. The rigorous theoretical analysis provided also validates these numerical findings. 

\paragraph*{Model.} To begin with, the phase-evolution of $N$ nonidentical Kuramoto oscillators~\cite{Kuramoto1984} in a simplicial complex under the impression of the rate-based learning of the 1-simplex and 2-simplex couplings is given by
\begin{eqnarray}\label{model}
	\dot\theta_{i} = \omega_{i} + \frac{\lambda_1}{\langle k^{[1]}\rangle}\sum_{j=1}^{N} A_{ij} \sin(\theta_{j}-\theta_{i}) + \frac{\lambda_2}{2!\langle k^{[2]}\rangle}\sum_{j,k=1}^{N} \nonumber\\ B_{ijk} \sin(2\theta_{j}-\theta_{k}-\theta_{i}),
\end{eqnarray}
where $\theta_i (\omega_i)\ (i{=}1,\hdots,N)$ denotes the instantaneous phases (intrinsic frequencies) of the $i$th oscillator in the simplicial complex and $\lambda_1$ and $\lambda_2$ are the coupling strengths of 1-simplex and 2-simplex interactions, respectively.
We conserve the global coupling of the 1-simplex and 2-simplex interactions in the complex by setting $\lambda = \lambda_1+\lambda_2$, irrespective of their topology assimilated in $A$ and $B$. This choice, for any given $\lambda$, allows us to maintain the dominance of one-type of simplex interaction over the other type  through a propensity parameter $p\in[0,1]$ such that $\lambda_1 = (1-p)\lambda = q\lambda$ and $\lambda_2 = p\lambda$. The number of edges or triangles in the complex a node is part of is defined as 1- or 2-simplex degrees, i.e., $k_i^{[1]} {=} \sum_{j=1}^N A_{ij}$ or $k_i^{[2]} {=} \frac{1}{2!}\sum_{j,k=1}^N B_{ijk}$, respectively, where $\langle k^{[1]}\rangle$ and $\langle k^{[2]}\rangle$ denote mean 1- and 2- simplex degrees, respectively. The 1- and 2- simplex coupling interactions are rescaled by the respective mean degrees so as to put the respective effective connectivities on an equal footing and assist $p$ in tuning the relative strengths of 1- and 2- simplex interactions:
\begin{eqnarray}\label{weight_adapt}
	\dot A_{ij} = \alpha\cos(\theta_j-\theta_i)-\mu A_{ij},\nonumber\\ 
	\dot B_{ijk}=\beta\cos(2\theta_{j}-\theta_{k}-\theta_{i})-\nu B_{ijk}.
\end{eqnarray}
We construct a 2-simplicial complex by identifying unique triangles and unique edges closing the triangles from a random 1-simplicial network.\ The collective phase evolution and weight adaptation of the adaptive 2-simplicial complex are then governed by Eqs.~(\ref{model}) and (\ref{weight_adapt}).\ To capture the formation of $m$ clusters in the network, we define an $m$-cluster order parameter $z_m{=}R_me^{i\Psi_m}{=}\frac{1}{N}\sum_{j=1}^{N}e^{im\theta_j}$ $(m{=}1,2)$, where $R_m$ and $\Psi_m$ are the amplitude and argument, respectively, of the $m$-cluster order parameter.\ Thus $R_1$ quantifies one-cluster synchronization whereas $R_2$ quantifies two-cluster synchronization~\cite{Daido1996}. 
\begin{figure}[t!]
	\centering
	\begin{tabular}{ccc}
	\includegraphics[height=4cm,width=4.2cm]{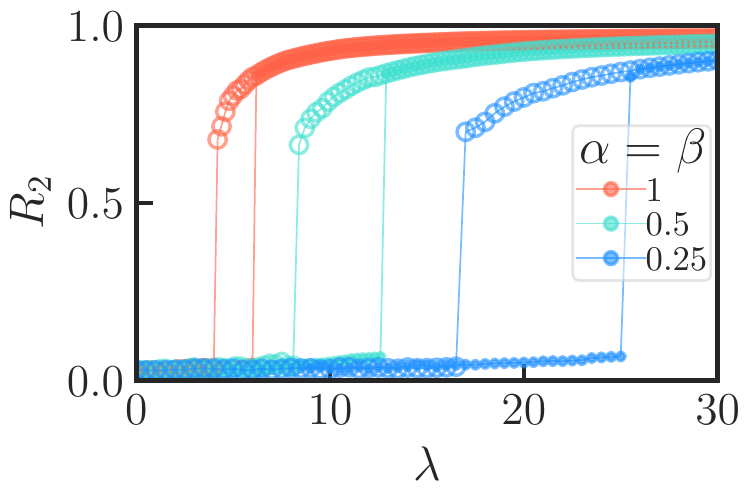}&
	\includegraphics[height=4cm,width=4.2cm]{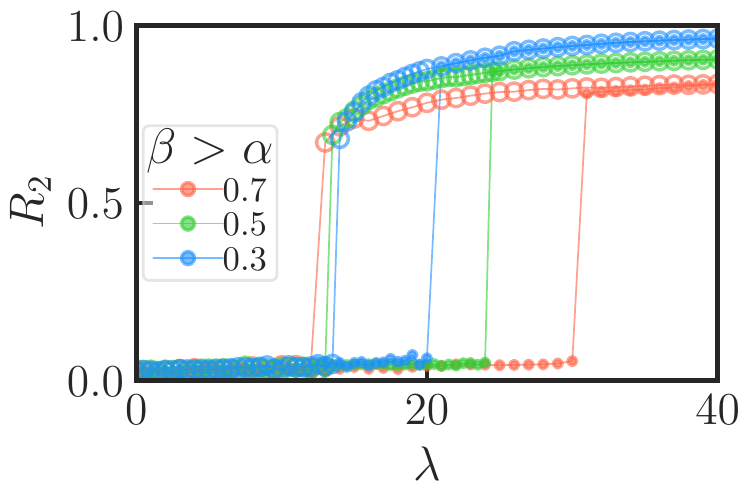}\\
	\end{tabular}{}
	\vspace{-0.5cm}
	\caption{(Color online) {Slow $\alpha{=}\beta$ and $\beta\textgreater\alpha$ prolong the onset of transition:} $R_2{-}\lambda$ synchronization profiles for the random 2-simplicial complex with adaptive couplings when $\alpha{=}\beta\le1$ and $\beta\textgreater\alpha$ $(=0.3)$.}
	\label{figure2}
\end{figure}

We numerically evolve Eqs.~(\ref{model}) and (\ref{weight_adapt}) to capture the microscopic dynamics of 1- and 2- simplex weights and the route to synchronization.\ All the results presented for the random 2-simplicial complex are for $N{=}10^3$, $\langle k^{[1]}\rangle{=}14$, and $\langle k^{[2]}\rangle{=}10$ with uniform randomly drawn natural frequencies $\omega_i{\sim} U(-\Delta, \Delta)$, where $\Delta{=}1$.\  All the initial 1-simplex (2-simplex) weights are equal and determined by $A_{ij}(0){=}1/L$ [$B_{ijk}(0){=}1/T$], where $L$ ($T$) are the number of 1-simplex (2-simplex) connections in the complex. At first, $\lambda$ is adiabatically increased until a large $\lambda$ and then adiabatically decreased until $\lambda=0$. The phase and weight dynamics [Eqs.~(\ref{model}) and (\ref{weight_adapt})] are then simultaneously simulated on the 2-simplicial complex and the order parameters are computed for each $\lambda$.

We now discuss the nature of the transition when the dyadic and triadic weights are static, i.e., only Eq.~(\ref{model}) is evolved, taking $A_{ij}{=}1$ and $B_{ijk}{=}1$ into account. Such static $A_{ij}$ and $B_{ijk}$ lead to a first-order (abrupt) transition to in-phase (single-cluster) synchronization $R_1$ with associated hysteresis, as shown in Fig.~\ref{figure1} (top row).\ The dyadic interactions are known to promote synchronization, while the triadic interactions do not. Hence when $p\textless0.5$, the dominating dyadic interactions quickly overcome the frustration induced by the triadic ones and lead to synchronization at lower values of $\lambda$ with rather reduced hysteresis width. The hysteresis is lost when dyadic interactions are much stronger than triadic ones for lower values of $p$. For $p\textgreater0.5$, the triadic interactions dominate, leading to significant frustration, eventually leading to abrupt synchronization at large $\lambda$ with broader hysteresis.\ For $p{=}1$, however $\lambda_f{\rightarrow}\infty$, as the only interaction-type existing among the oscillators, is the triadic one that does not lead to synchronization for any $\lambda>0$ ~\cite{Skardal2019}.

Nevertheless, incorporating adaptive 1-simplex and 2-simplex couplings [Eq.~(\ref{weight_adapt})] and the phase evolution [Eq.~(\ref{model})] results in the finding that the in-phase synchronization $R_1$ is completely subsided; instead, a first-order antiphase (two-cluster) synchronization $R_2$ emanates with an increase in $\lambda$ [see Fig.~\ref{figure1} (bottom row)]. Although the nature of the abrupt $R_2$ transition for different values of $p$ is analogous to that of the abrupt $R_1$ transition of the static case, the onset of the abrupt $R_2$ occurs at larger values of $\lambda$ than of $R_1$.\ Not that the adaptive pure 2-simplex couplings (when $p{=}1$) do not lead to either in-phase or antiphase synchronization with the increase in $\lambda$~\cite{Kachhvah2022}, whereas the adaptive pure 1-simplex couplings (when $p{=}0$) lead to a second-order anti-phase synchronization while in-phase synchronization does not occur~\cite{Kachhvah2022}. One remarkable feature of the emergent abrupt antiphase synchronization and desynchronization is that their respective onsets are entirely manageable through the learning parameters $\alpha, \beta, \mu$, and $\nu$. Further, the impact of dyadic and triadic learning rates on the characteristics of the $R_2$ transition is illustrated in Fig.~\ref{figure2}, which shows that the slower learning rates $\alpha{=}\beta$ delay the outset of the abrupt transition to a higher $\lambda_f$.\ In addition, $\beta\textgreater\alpha$ also triggers the abrupt transition at a higher $\lambda_f$.

\begin{figure}[t!]
	\centering
	\includegraphics[height=9cm,width=8.5cm]{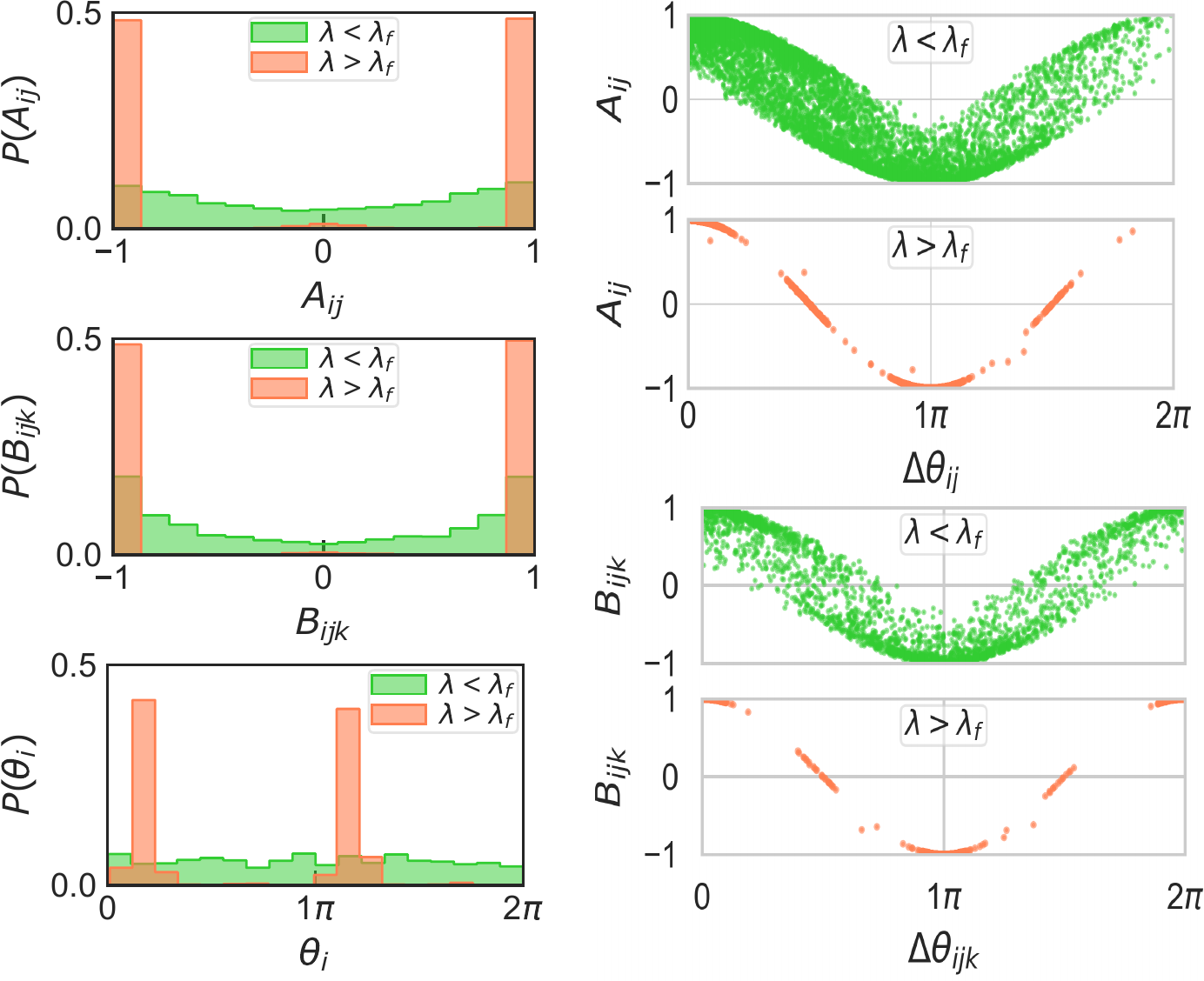}\\
	\vspace{-0.4cm}
	\caption{(Color online) {\em Steady state attributes:} Distributions $P(A_{ij})$, $P(B_{ijk})$ and $P(\theta_i)$ of the stationary $A_{ij}$, $B_{ijk}$ and $\theta_i$, respectively, and $A_{ij}$ and $B_{ijk}$ plotted against $\Delta\theta_{ij}$ and $\Delta\theta_{ijk}$, respectively.\ All the results are carried out for the random 2-simplicial complex with $\alpha{=}\beta{=}1$ and $\mu{=}\nu{=}1$.}
	\label{figure3}
\end{figure}

Further, we shed light on the distribution of stationary phases and adaptive dyadic and triadic weights in the incoherent and coherent states (see the left panels of Fig.~\ref{figure3}). In the coherent state for $\lambda\textgreater\lambda_f$, the stationary $A_{ij}$ ($B_{ijk}$) are segregated into two clusters. Hence the distribution $P(A_{ij})$ [$P(B_{ijk})$] manifests bimodal peaks at $-\alpha/\mu$ (-$\beta/\nu$) and $\alpha/\mu$ ($\beta/\nu$), with a few $A_{ij}$ and $B_{ijk}$ settling on approximately $0$. The corresponding phases are also set apart into bimodal peaks at a difference of $\pi$, resulting in $P(\theta_{i})$ exhibiting antiphase clusters. Nevertheless in the incoherent state for $\lambda\textless\lambda_f$, $P(A_{ij})$ [$P(B_{ijk})$] follows a $\beta$ distribution with peaks at $-\alpha/\mu$ ($-\beta/\nu$) and $\alpha/\mu$ ($\beta/\nu$) and dips at $0$ ($0$).\ Moreover, $\dot A_{ij}{=}0$ and $\dot B_{ijk}{=}0$ yield the dyadic and triadic stationary weights 
\begin{eqnarray}\label{steadyWeights}
	A_{ij} = \frac{\alpha}{\mu} \cos(\Delta\theta_{ij}),\quad B_{ijk} = \frac{\beta}{\nu}\cos(\Delta\theta_{ijk}),
\end{eqnarray}
where $\Delta\theta_{ij}{=}(\theta_j{-}\theta_i)$ and $\Delta\theta_{ijk}{=}(2\theta_j{-}\theta_k{-}\theta_i)$.\ Equations~(\ref{steadyWeights}) corroborate the numerical revelations of Fig.~\ref{figure3}. In the coherent state, the steady-state extrema $A_{ij}{\rightarrow}\pm\alpha/\mu$ and $B_{ijk}{\rightarrow\pm}\beta/\nu$ correspond to $\Delta\theta_{ij}{\rightarrow}0,\pi$ and $\Delta\theta_{ijk}{\rightarrow}0,\pi$, respectively.\ Also the steady-state $A_{ij}{\rightarrow}0$ and $B_{ijk}{\rightarrow}0$ are associated with $\Delta\theta_{ij}{\rightarrow}\pi/2,3\pi/2$ and $\Delta\theta_{ijk}{\rightarrow}\pi/2,3\pi/2$, respectively. 
Nonetheless, in the incoherent state, the uniformly distributed stationary phases require $\Delta\theta_{ij}$ and $\Delta\theta_{ijk}$ to draw the phases from the full range $[0, 2\pi)$. Thereby, the stationary $A_{ij}$ and $B_{ijk}$ acquire the weights from the full intervals $[-\alpha/\mu,\alpha/\mu]$ and $[-\beta/\nu,\beta/\nu]$, respectively.

\paragraph*{Ott-Antonsen reduction:}
To seek analytical insight into the underlying higher-order dynamics, we turn our focus to an all-to-all connected 2-simplicial complex modeled as
\begin{eqnarray}\label{GCmodel}
	\dot\theta_{i} = \omega_{i} + \frac{q\lambda}N\sum_{j=1}^{N} A_{ij} \sin(\theta_{j}-\theta_{i}) + \frac{p\lambda}{N^2}\sum_{j,k=1}^{N}B_{ijk}\nonumber \\ \sin(2\theta_{j}-\theta_{k}-\theta_{i}).
\end{eqnarray}
As per Ott and Antonsen~\cite{Ott2009}, the long-time evolution of the order parameter for a system involving adaptive coupling obeys the single differential equation achieved using the Ott-Antonsen ansatz~\cite{Ott2008} as the precise time dependence of the adaptive coupling would not matter in this analytical treatment. 
Hence we employ Ott-Antonsen dimensionality reduction to the steady-state collective dynamics of Eqs.~(\ref{weight_adapt}) and (\ref{GCmodel}). The steady-state collective dynamics can only be achieved when the phases and dyadic and triadic weights simultaneously achieve their respective steady states. 
The evolution of phases can be described, after plugging into the steady-state expressions for $A_{ij}$ and $B_{ijk}$, as
\begin{eqnarray}\label{eqn5}
	\dot\theta_{i} = \omega_{i} + \frac{aq\lambda}{2N} \sum_{j=1}^{N} \sin(2\theta_j-2\theta_{i}) + \frac{bp\lambda}{2N^2} \sum_{j,k=1}^{N}\nonumber\\\sin(4\theta_j-2\theta_k-2\theta_{i}),
\end{eqnarray}
where $a={\alpha}/{\mu}$ and $b={\beta}/{\nu}$. Note that the footprints of both $A_{ij}$ and $B_{ijk}$ are assimilated into Eq.~(\ref{eqn5}) in the form of higher modes of phases in the attractive dyadic and triadic couplings, respectively. The phase evolution can be reexpressed further in terms of the $m$-cluster order parameters
\begin{eqnarray}\label{selfconst}
	\dot\theta_i=\omega_i+\frac{1}{4i}[He^{-2i\theta}-H^* e^{2i\theta}],\nonumber\\ 
	H = aq\lambda z_2 + bp\lambda z_2^*z_4. 
\end{eqnarray}
Considering the system in the continuum limit $N{\rightarrow}\infty$, the collective state of the oscillators at a time $t$ can be delineated by a continuous density function $\rho(\theta,\omega,t)$ such that $\rho(\theta,\omega,t)\mathrm{d\theta d\omega}$ denotes the fraction of oscillators with their phases and intrinsic frequencies lying in the ranges of $[\theta,\theta+d\theta]$ and $[\omega,\omega+d\omega]$, respectively.\ In addition, the density function $\rho(\theta,\omega,t)$ satisfies the normalization condition $\int_{0}^{2\pi}\rho(\theta,\omega,t)\mathrm{d\theta}{=}1$ and the continuity equation $\partial_t \rho(\theta,\omega,t)+ \partial_{\theta} [\rho(\theta,\omega,t)\ v(\theta,\omega,t)]{=}0$ as the number of oscillators remains conserved.\ Also, the $m$-cluster order parameter can be expressed as $z_m{=}\int\int \mathrm{d\omega d\theta} e^{im\theta} \rho(\theta,\omega,t)$. Since $\rho(\theta,\omega,t)$ is a $2\pi$-periodic function with respect to $\theta$, it can be expressed as a Fourier expansion of the form
\begin{eqnarray}\label{funcFourier}
	\rho(\theta,\omega,t)=\frac{1}{2\pi}\bigg[1+\bigg\{\sum_{n=1}^{\infty}f_n(\omega,t)e^{in\theta} + c.c.\bigg\}\bigg],
\end{eqnarray}
where c.c. stands for the complex conjugate of the preceding terms.\
Ott and Antonsen pointed out that all Fourier coefficients can be classified into Poisson kernels of the form $f_n(\omega,t)=[f(\omega,t)]^n$, where $|f(\omega,t)|\ll1$ is necessary for the convergence of the series.
After plugging into the expressions for $v=\dot\theta$ [Eq.~(\ref{selfconst})] and $f(\theta,\omega,t)$ [Eq.~(\ref{funcFourier})], all Fourier modes then reduce to the same constraint for $f$, satisfying the single complex-valued differential equation
\begin{align}\label{OAeq}
	&\frac{\partial f^2} {\partial t} + 2i\omega f^2 + \frac{1}{2}[H f^4 - H^*]=0,\\
	&z_m=\mathcal{G}f^{*m}=\int_{-\infty}^{\infty}\mathrm{d\omega g(\omega)}f^{*m}(\omega,t),\ m{=}2,4,
\end{align}
where the integral operator $\mathcal{G}\equiv\int_{-\infty}^{\infty}\mathrm{d\omega g(\omega)}$.

\begin{figure}[t!]
	\centering
	\begin{tabular}{cc}
	\includegraphics[height=4.15cm,width=4.5cm]{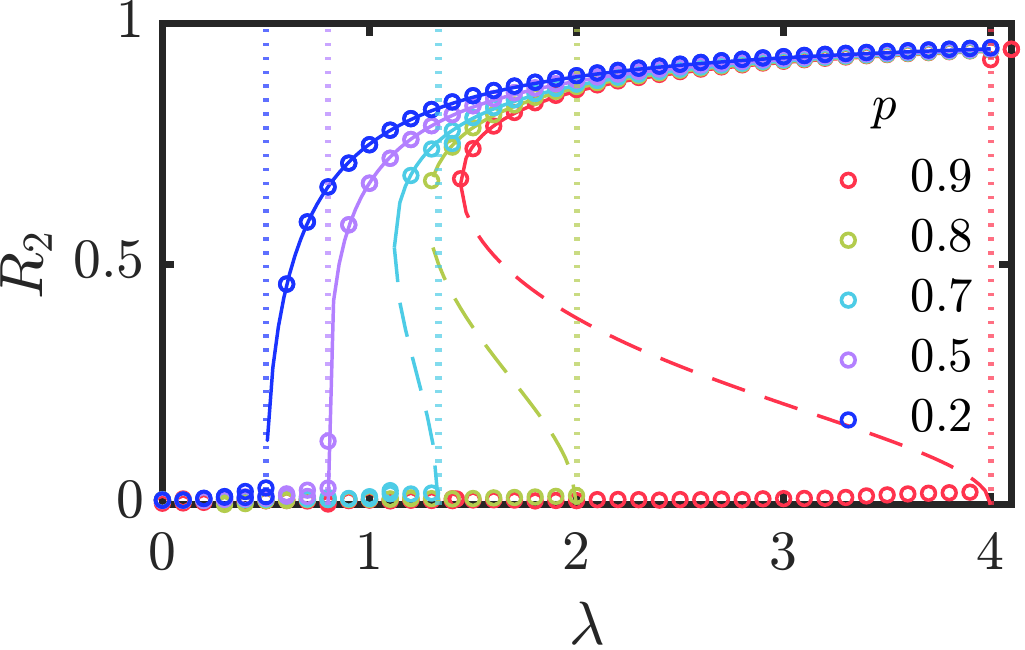}&
	\includegraphics[height=4cm,width=3.8cm]{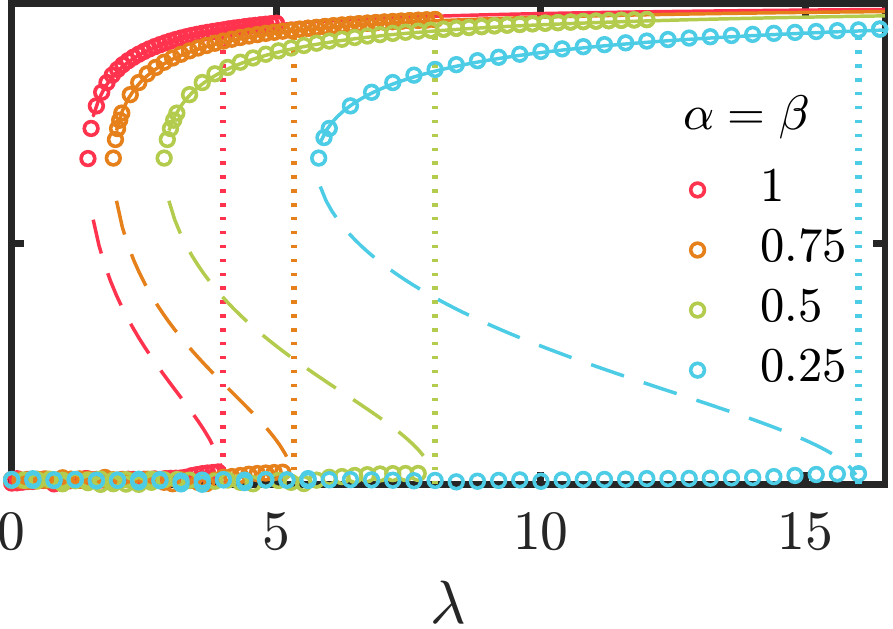}\\
	\end{tabular}{}
	\vspace{-0.5cm}
	\caption{(Color online) {Analytical vs. numerical corroboration of $R_2$:} $R_2{-}\lambda$ profiles of the all-to-all connected 2-simplicial complex simulated for Lorentzian $g(\omega)$ with $\Delta{=}0.1$, $N{=}10^4$, $\mu{=}1$, $\nu{=}1$. Shown on the left are different values of $p$ when $\alpha{=}\beta{=}1$ and on the right are different values of $\alpha{=}\beta$ when $p{=}0.9$. The solid and dashed lines are the respective theoretical traces of $R_2^+$ (stable) and $R_2^-$ (unstable) solutions obtained using Eq.~(\ref{OAopara}). The dotted lines are analytical predictions [Eq.~(\ref{fcrit})] of $\lambda_f$.}
	\label{figure4}
\end{figure}
\paragraph*{Stability of the incoherent state:} The trivial solution $f(\omega,t){=}0$ always exists for Eq.~(\ref{OAeq}), which corresponds to an incoherent state $\rho(\theta,\omega,t){=}\frac{1}{2\pi}$ in Eq.~(\ref{funcFourier}). 
Linearizing Eq.~(\ref{OAeq}) around $f(\omega,t){=}0$, we obtain the following linear equation for the perturbed density $\eta(\omega,t)$:
\begin{eqnarray}\label{evo_pert}
	\frac{\partial\eta^2}{\partial t} + 2 i \omega \eta^2 = \frac{aq\lambda}{2} \mathcal{G}\eta^2.
\end{eqnarray}
Let $\gamma$ be the eigenvalues of Eq.~(\ref{evo_pert}) such that $\eta(\omega,t)=\eta_0(\omega)\ e^{\gamma t}$.\ Then employing the integral operator $\mathcal{G}$ to both sides of Eq.~(\ref{evo_pert}) reduces it to
\begin{eqnarray}\label{coup_ev}
	\frac{1}{\lambda}=\frac{aq}{4}\int_{-\infty}^{\infty}d\omega\ \frac{g(\omega)}{\gamma+i \omega}.
\end{eqnarray}
Since $Re[\gamma]{=}0$ at the critical coupling strength $\lambda{=}\lambda_f$, the incoherent state loses stability for $\gamma{=}0+\epsilon'+iy$, where $0<\epsilon'\ll1$. Equation~(\ref{coup_ev}) now reads 
\begin{eqnarray}\label{coup_evp}
	\frac{1}{\lambda}=\frac{aq}{4} \lim_{\epsilon'\rightarrow0} \int_{-\infty}^{\infty}d\omega\ \frac{g(\omega)}{\epsilon'+i\omega+iy}.
\end{eqnarray}
Solving Eq.~(\ref{coup_evp}) for $g(\omega){=}\frac{\Delta}{\pi[\omega^2+\Delta^2]}$ results in 
\begin{eqnarray}\label{fcrit}
	\lambda_f=\frac{4\Delta}{aq}=\frac{4\mu\Delta}{\alpha (1-p)}.
\end{eqnarray}
Equation~(\ref{fcrit}) reveals that the transition to synchronization is solely caused by the presence of dyadic interactions through the parameter $p$ and dyadic rates $\alpha$ and $\mu$ and the triadic interactions do not play any role in the onset of synchronization.\ Also, $p{=}1$ yields $\lambda_f{\rightarrow}\infty$, i.e., the incoherence does not lose stability for any $\lambda>0$ ~\cite{Kachhvah2022}.\\

\paragraph*{Solution of coherence:} 
The expression for order parameter $R_2$ can be worked out for a Lorentzian distribution with mean $\omega_0$ and half-width $\Delta$, i.e., $g(\omega){=}\frac{\Delta}{\pi[(\omega-\omega_0)^2+\Delta^2]}$.\ The order parameter in Eq.~(\ref{OAeq}) can be derived using Cauchy’s residue theorem by closing the contour to an infinite-radius semi-circle in the negative-half complex $\omega$ plane, resulting in 
	$z_2{=}f^{*2}(\omega_0{-}i\Delta,t)$ and $z_4{=}f^{*4}(\omega_0{-}i\Delta,t){=}z_2^2$.\ Next, assessing Eq.~(\ref{OAeq}) at $\omega{=}\omega_0{-}i\Delta$ and then taking the complex conjugate, we obtain
\begin{eqnarray}
	2\dot z_2 - 4i\omega_0z_2 + 4\Delta z_2 + \lambda z_2^2 [a q z_2^* + b p z_2 z_4^*]\nonumber\\ -\lambda[a q z_2 + b p z_2^* z_4] = 0.
\end{eqnarray}
Next, inserting $z_2=R_2e^{i\Psi_2}$ and then equating real and imaginary parts on both sides of the equation gives
\begin{eqnarray}\label{OAfinal}
	&2\dot R_2 + 4\Delta R_2 + \lambda R_2 (R_2^2-1) (a q + b p R_2^2)=0,\\
	&\dot\Psi_2=2\omega_0.
\end{eqnarray}
Hence the dynamics of $R_2$ and $\Psi_2$ are decoupled;\ $\Psi_2$ constantly evolves and is equal to twice the mean of $g(\omega)$.\ The steady-state evolution of Eq.~(\ref{OAfinal}) yields an equation that is cubic in $R_2$.\ Thus, $R_2{=}0$ is always an equilibrium whose stability is not affected by the presence of higher-order interaction.\ The nonlinear terms stemming from higher-order interaction in
\begin{eqnarray}\label{steadyState}
	\frac{4\Delta}{\lambda} = -bp R_2^4 + (bp - aq) R_2^2 + aq 
\end{eqnarray}
may give rise to one or two synchronous solutions with the positive roots for $R_2$
\begin{eqnarray}\label{OAopara}
	R_2^{\pm}{=}\sqrt{\frac{bp-aq\pm\sqrt{\big(bp-aq\big)^2+4bp\big(aq-\frac{4\Delta}{\lambda}\big)}}{2bp}},
\end{eqnarray}
where $R_2^+$ ($R_2^-$) represents a stable (an unstable) branch of the synchronous state. The validation of analytical predictions for the order parameter $R_2$ [Eq.~(\ref{OAopara})] with its numerical estimations for different sets of parameters is presented in Fig.~\ref{figure4}.

\begin{figure}[t!]
	\centering
	\begin{tabular}{cc}
	\includegraphics[height=3.8cm,width=4.3cm]{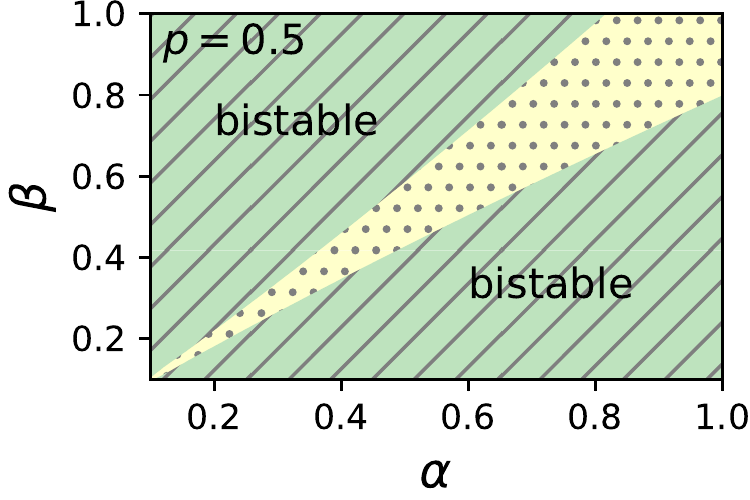}&
	\includegraphics[height=3.8cm,width=4.3cm]{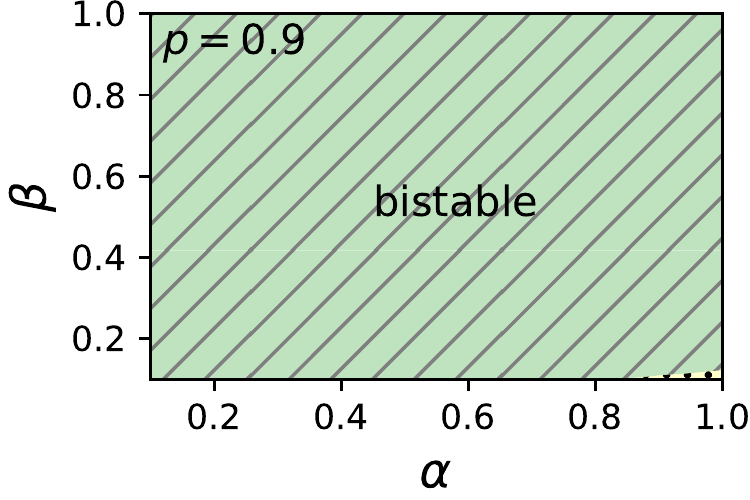}\\
	\includegraphics[height=3.8cm,width=4.2cm]{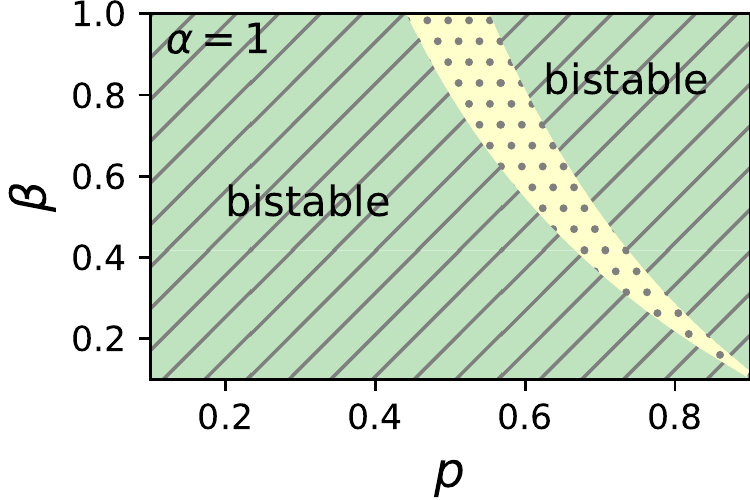}&
	\includegraphics[height=3.8cm,width=4.2cm]{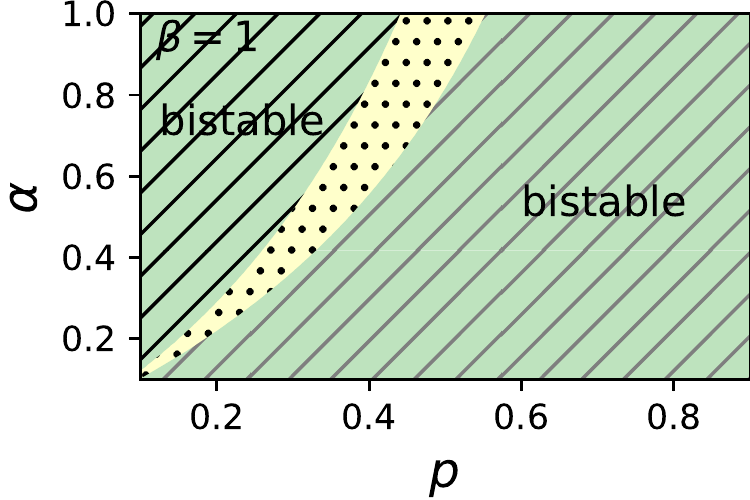}\\
	\end{tabular}{}
	\vspace{-0.5cm}
	\caption{(Color online) {Bistability domains:} phase diagrams in $\alpha-\beta$, $p-\alpha$ and $p-\beta$ planes depicting the regions of bistability for the all-to-all connected 2-simplicial complex simulated for Lorentzian $g(\omega)$, with $\Delta{=}0.1$, $N{=}10^3$, $\mu{=}1$, and $\nu{=}1$. The dotted region manifests nonbistability.}
	\label{figure5}
\end{figure}
In the case of a forward transition, in the incoherent state $R_2{=}0$ until $\lambda{=}\lambda_f{=}\frac{4\Delta}{aq}$ is reached. At $\lambda{=}\lambda_f$, $R_2$ abruptly jumps to $R_2^+(\lambda_f){=}\sqrt{1-\frac{aq}{bp}}$ [while $R_2^-(\lambda_f){=0}$] and the incoherent state ($R_2{=}0$) loses its stability through subcritical pitchfork bifurcation.\ Nevertheless, for the set of parameters $\{a, b, p\}$ for which $R_2^+(\lambda_f){=}\sqrt{1-\frac{aq}{bp}}{=}0$ at $\lambda{=}\lambda_f$, the incoherent state ($R_2{=}0$) loses its stability through supercritical pitchfork bifurcation and the transition to synchronization takes place via a second-order route.

In the case of a backward transition, $R_2^-$ (saddle point) and $R_2^+$ (node point) exist in the hysteresis region. As soon as the backward critical coupling strength $\lambda{=}\lambda_b$ is reached, $R_2^+$ and $R_2^-$ collide and annihilate each other through saddle-node bifurcation. Thus the stability of the coherent state is totally destroyed and the only remaining solution is $R_2{=}0$. Hence the constraint $d\lambda/d R_2{=}0$ is satisfied at $\lambda{=}\lambda_b$, which leads to $R_2(\lambda_b)=\sqrt{\frac{1}{2}(1 - \frac{aq}{bp})}$ from Eq.~(\ref{steadyState}). Substituting the value of $R_2$ back into Eq.~(\ref{steadyState}) gives
\begin{eqnarray}\label{bcrit}
	\lambda_b=\frac{16\Delta b p}{\big[aq+bp\big]^2}=\frac{16\Delta p\beta\nu\mu^2}{\big[\alpha\nu(1-p)+\beta \mu p\big]^2}.
\end{eqnarray}
For that matter, the outset of abrupt desynchronization is characterized by both the dyadic and triadic learning rates.

In Fig.~\ref{figure5} we provide a broad picture of the regions of bistability and nonbistability stretched over $\alpha{-}\beta$, $p{-}\alpha$ and $p{-}\beta$ planes~\cite{SM}.\ The regions illustrated by the slanted green lines represent the bistability region. Note that $aq{\ne}bp$ and $aq\textless bp$ are necessary for the existence of bistable solutions sporting a hysteresis. The yellow dotted region depicts the nonbistable region corresponding to $aq{=}bp$.~\footnote{The nonbistable region also includes the set of parameters yielding hysteresis width tending to zero, i.e., $|\lambda_f-\lambda_b|\textless0.01$.} For $aq{=}bp$, Eqs.~(\ref{fcrit}) and (\ref{bcrit}) furnish $\lambda_f{=}\lambda_b$, $R_2^+(\lambda_f){=}0$, and $R_2(\lambda_b){=}0$, which conform to a second-order transition to synchronization.
\\

\paragraph*{Conclusion}
In this work, the nature of the transition to synchronization was explored on 2-simplicial complexes where the triadic couplings and the downward closing dyadic couplings evolve in time according to the respective rate-based plasticity inspired by the Hebbian learning rule.\ Strikingly, such coevolving dyadic and triadic couplings completely subside the single-cluster synchronization and instead trigger two-cluster synchronization in simplicial complexes. It was revealed that the onset of antiphase synchronization only depends on the dyadic interaction (learning rate) and the higher-order interaction has no role to play. On the other hand, both dyadic and triadic interactions (learning rates) affect the onset of antiphase desynchronization.\ Further, the numerical findings related to the antiphase order parameter and the forward and backward critical transition points have been validated with the respective analytical predictions by employing the Ott-Antonsen ansatz. It was also shown that the steady dyadic (triadic) weights in the synchronous state form two clusters of equal and opposite magnitudes along the lines of the oscillators forming the antiphase clusters.

The simplicial structures involving simultaneous adaptation of pairwise and higher-order interactions would help elucidate the underlying mechanism of cluster formation in the brain's functional networks, such as, antiphase patterns in the cortical neural network.

\begin{acknowledgments}
SJ acknowledges support through Government of India, Department of Science and Technology (DST) POWER Grant No. SERB/F/9035/2021-2022.
\end{acknowledgments}


\begin{thebibliography}{99}

	\bibitem{Petri2014} G. Petri, P. Expert, F. Turkheimer, R. Carhart-Harris, D. Nutt, P. J. Hellyer and F. Vaccarino, {Homological scaffolds of brain functional networks}, {\em J. R. Soc. Interface} {\bf 11}, 20140873 (2014)

	\bibitem{Benson2018} A. R. Benson, R. Abebe, M. T. Schaub, A. Jadbabaie and J. Kleinberg, {Simplicial closure and higher-order link prediction}, {\em Proc. Natl. Acad. Sci. USA} {\bf 115}, E11221 (2018)
	
	\bibitem{Iacopini2019} I. Iacopini, G. Petri, A. Barrat, and V. Latora, {Simplicial models of social contagion}, {\em Nat. Commun.} {\bf 10}, 2485 (2019) 
	
	\bibitem{Matamalas2020}  J. T. Matamalas, S. Gómez and A. Arenas, {Abrupt phase transition of epidemic spreading in simplicial complexes}, {\em Phys. Rev. Research} {\bf 2}, 012049(R) (2020)

	\bibitem{Salnikov2018} V. Salnikov, D. Cassese, and R. Lambiotte, {Simplicial complexes and complex systems}, {\em Eur. J. Phys.} {\bf 40}, 014001 (2019)
	
	\bibitem{Whitehead1939} J. H. C. Whitehead, {Simplicial spaces, nuclei and m-groups}, {\em Proc. London Math. Soc.} {\bf s2-45}, 243 (1939)
	
	\bibitem{Tanaka2011} T. Tanaka and T. Aoyagi, {Multistable Attractors in a Network of Phase Oscillators with Three-Body Interactions}, {\em Phys. Rev. Lett.} {\bf 106}, 224101 (2011)
	
	\bibitem{Skardal2019} P. S. Skardal and A. Arenas, {Abrupt Desynchronization and Extensive Multistability in Globally Coupled Oscillator Simplexes}, {\em Phys. Rev. Lett.} {\bf 122}, 248301 (2019) 
	
	\bibitem{Xu2020} C. Xu, X. Wang and P. S. Skardal, {Bifurcation analysis and structural stability of simplicial oscillator populations}, {\em Phys. Rev. Research} {\bf 2}, 023281 (2020)
	
	\bibitem{Millan2020} A. P. Mill{\'a}n, J. J. Torres and G. Bianconi, {Explosive Higher-Order Kuramoto Dynamics on Simplicial Complexes}, {\em Phys. Rev. Lett.} {\bf 124}, 218301 (2020)
	
	\bibitem{Skardal2020} P. S. Skardal and A. Arenas, {Higher order interactions in complex networks of phase oscillators promote abrupt synchronization switching}, {\em Commun. Phys.} {\bf 3}, 218 (2020)
	
	\bibitem{Lucas2020} M. Lucas, G. Cencetti and F. Battiston, {Multiorder Laplacian for synchronization in higher-order networks}, {\em Phys. Rev. Research} {\bf 2}, 033410 (2020)
	
	\bibitem{Battiston2020} F. Battiston, G. Cencetti, I. Iacopini, V. Latora, M. Lucas, A. Patania, Jean-Gabriel Young and G. Petri, {Networks beyond pairwise interactions: Structure and dynamics}, {\em Phys. Rep.} {\bf 874}, 1 (2020)
	
	\bibitem{Ghorbanchian2021} R. Ghorbanchian, J. G. Restrepo, J. J. Torres and G. Bianconi, {Higher-order simplicial synchronization of coupled topological signals}, {\em Commun. Phys.} {\bf 4}, 120 (2021)
	
	\bibitem{Xu2021} C. Xu and P. S. Skardal, {Spectrum of extensive multiclusters in the Kuramoto model with higher-order interactions}, {\em Phys. Rev. Research} {\bf 3}, 013013 (2021)
	
	\bibitem{Chutani2021} M. Chutani, B. Tadi{\'c}, and N Gupte, {Hysteresis and synchronization processes of Kuramoto oscillators on high-dimensional simplicial complexes with competing simplex-encoded couplings}, {\em Phys. Rev. E} {\bf104}, 034206 (2021)
	
	\bibitem{Sun2021} H. Sun and G. Bianconi, {Higher-order percolation processes on multiplex hypergraphs}, {\em Phys. Rev. E} {\bf 104}, 034306 (2021)
	
	\bibitem{Kuehn2021} C. Kuehn and C. Bick, {A universal route to explosive phenomena}, {\em Sci. Adv.} {\bf 7}, eabe3824 (2021)
	
	\bibitem{Dabaghian2012} Y. Dabaghian, F. M{\'e}moli, L. Frank, G. Carlsson, {A Topological paradigm for Hippocampal spatial map formation using persistent homology}, {\em PLoS Comput. Biol.} {\bf 8}, e1002581 (2012)
	
	\bibitem{Shimizu2000} E. Shimizu, Y. P. Tang, C. Rampon and J. Z. Tsien, {NMDA receptor-dependent synaptic reinforcement as a crucial process for memory consolidation}, {\em Science} {\bf 290}, 1170 (2000)
	
	\bibitem{Abbott2000}L. F. Abbott and S. B. Nelson, {Synaptic plasticity: Taming the beast}, {\em Nat. Neurosci.} {\bf 3}, 1178 (2000)
	
	\bibitem{Hebb1949} D. O. Hebb, {\em The Organization of Behavior} (Wiley, New York, 1949) 
	
	\bibitem{Markram1997} H. Markram, J. L{\"u}bke, M. Frotscher and B. Sakmann, {Regulation of synaptic efficacy by coincidence of postsynaptic APs and EPSPs}, {\em Science} {\bf 275}, 213 (1997)
	
	\bibitem{Zhang1998} L. I. Zhang, H. W. Tao, C. E. Holt, W. A. Harris and M. Poo, {A critical window for cooperation and competition among developing retinotectal synapses}, {\em Nature (London)} {\bf 395}, 37 (1998)
	
	\bibitem{Zhigulin2003} V. P. Zhigulin, M. I. Rabinovich, R. Huerta, and H. D. I. Abarbanel, {Robustness and enhancement of neural synchronization by activity-dependent coupling}, {\em Phys. Rev. E} {\bf 67}, 021901 (2003)
	
	\bibitem{Knoblauch2012} A. Knoblauch, F. Hauser, M-O Gewaltig, E. K{\"o}rner, and G. Palm, {Does spike-timing-dependent synaptic plasticity couple or decouple neurons firing in synchrony?}, {\em Front. Comput. Neurosci.} {\bf 6}, 55 (2012)
	
	\bibitem{Zirkle2020} J. Zirkle, and L. L. Rubchinsky, {Spike-timing dependent plasticity effect on the temporal patterning of neural synchronization}, {\em Front. Comput. Neurosci.} {\bf 14}, 52 (2020)
	
	\bibitem{Niyogi2009} R. K. Niyogi and L. Q. English, {Learning-rate-dependent clustering and self-development in a network of coupled phase oscillators}, {\em Phys. Rev. E} {\bf 80}, 066213 (2009)
	
	\bibitem{Aoki2009} T. Aoki and T. Aoyagi, {Co-evolution of Phases and Connection Strengths in a Network of Phase Oscillators}, {\em Phys. Rev. Lett.}  {\bf 102}, 034101 (2009)
 	
 	\bibitem{Gutierrez2011} R. Guti\'errez, A. Amann, S. Assenza, J. G\'omez-Garde\~nes, V. Latora, and S. Boccaletti, {Emerging Meso- and Macroscales from Synchronization of Adaptive Networks}, {\em Phys. Rev. Lett.} {\bf 107}, 234103 (2011)
	
	\bibitem{Pitsik2018} E. Pitsik, V. Makarov, D. Kirsanov, N. Frolov, M. Goremyko, X. Li, Z. Wang, A. Hramov and S. Boccaletti, {Inter-layer competition in adaptive multiplex network}, {\em New J. Phys.} {\bf 20}, 075004 (2018)
	
	\bibitem{Khanra2022} P. Khanra, S. Ghosh, K. Alfaro-Bittner, P. Kundu, S. Boccaletti, C. Hens, and P. Pal, {Identifying symmetries and predicting cluster synchronization in complex networks}, {\em Chaos Soliton. Fract.} {\bf 155}, 111703 (2022)
	
	\bibitem{Shepelev2021} I. Shepelev, A. Bukh, G. Strelkova and V. Anishchenko, {Anti-phase relay synchronization of wave structures in a heterogeneous multiplex network of 2D lattices}, {\em Chaos Soliton. Fract.} {\bf 143}, 110545 (2021)
	
	\bibitem{AGaytan2018} V. Avalos-Gayt\'an, J. A. Almendral, I. Leyva, F. Battiston, V. Nicosia, V. Latora and S. Boccaletti, {Emergent explosive synchronization in adaptive complex networks}, {\em Phys. Rev. E} {\bf 97}, 042301 (2018)
  
	\bibitem{Kachhvah2020} A. D. Kachhvah, X. Dai, S. Boccaletti, and S Jalan, {Interlayer Hebbian plasticity induces first-order transition in multiplex networks}, {\em New J. Phys.} {\bf 22}, 122001 (2020)
	
	\bibitem{Frolov2021} N. Frolov, and A. Hramov, {Coexistence of interdependence and competition in adaptive multilayer network}, {\em Chaos Soliton. Fract.} {\bf 147}, 110955 (2021)
	
	\bibitem{Berner2021} R. Berner, S. Vock, E. Sch\"oll, and S. Yanchuk, {Desynchronization Transitions in Adaptive Networks}, {\em Phys. Rev. Lett.} {\bf 126}, 028301 (2021)
	
	\bibitem{Majhi2022} S. Majhi, M. Perc, and D. Ghosh, {Dynamics on higher-order networks: A review}, {\em J. R. Soc. Interface} {\bf 19}, 20220043 (2022)
	
	\bibitem{Lodi2020} M. Lodi, F. D. Rossa, F. Sorrentino, and M. Storace, {Analyzing synchronized clusters in neuron networks}, {\em Sci. Rep.}  {\bf 10}, 16336 (2020)
	
	\bibitem{Pecora2014} L. M. Pecora, F. Sorrentino, A. M. Hagerstrom, T. E. Murphy, and R. Roy, {Cluster synchronization and isolated desynchronization in complex networks with symmetries}, {\em Nat. Commun.}  {\bf 5}, 4079 (2014)
	
	\bibitem{Schaub2016} M. T. Schaub, N. O’Clery, Y. N. Billeh, J-C Delvenne, R. Lambiotte, and M. Barahona, {Graph partitions and cluster synchronization in networks of oscillators}, {\em Chaos}  {\bf 26}, 094821 (2016)
	
	\bibitem{Hemelrijk2012} C. K. Hemelrijk, and H. Hildenbrandt, {Schools of fish and flocks of birds: Their shape and internal structure by self-organization}, {\em Interface Focus}  {\bf 2}, 726 (2012)
	
	\bibitem{Kuramoto1984} Y. Kuramoto, {Self-entrainment of a population of coupled non-linear oscillators}, in {\em International Symposium on Mathematical Problems in Theoretical Physics}, {edited by H. Araki}, {Lecture Notes in Physics Vol. 39} (Springer, Berlin, 1975)
	
	\bibitem{Daido1996} H. Daido, {Multibranch Entrainment and Scaling in Large Populations of Coupled Oscillators}, {\em Phys. Rev. Lett.}  {\bf 77}, 1406 (1996)
	
	\bibitem{Kachhvah2022} A. D. Kachhvah, and S. Jalan, {Hebbian plasticity rules abrupt desynchronization in pure simplicial complexes}, {\em New J. Phys.} {\bf 24}, 052002 (2022)
	
	\bibitem{Ott2009} E. Ott, and T. M. Antonsen, {Long time evolution of phase oscillator systems}, {\em Chaos}  {\bf 19}, 023117 (2009)
	
	\bibitem{Ott2008} E. Ott and T. M. Antonsen, {Low dimensional behavior of large systems of globally coupled oscillators}, {\em Chaos} {\bf 18}, 037113 (2008)
	
	\bibitem{SM} See Supplemental Material at \url{} for details on the stationary dyadic and triadic weights, and alternative bistability phase diagrams. 
		
\end{thebibliography}
\end{document}

\title{Supplemental Material: First-order route to antiphase clustering in adaptive simplicial complexes}

\author{Ajay Deep Kachhvah, and Sarika Jalan}
\affiliation{Complex Systems Lab, Department of Physics, Indian Institute of Technology Indore, Khandwa Road, Simrol, Indore-453552, India}

\maketitle

\subsection{Stationary dyadic and triadic weights as a function of coupling strength}
\begin{figure}[h!]
	\centering
	\begin{tabular}{cc}
	\includegraphics[height=4.5cm,width=6cm]{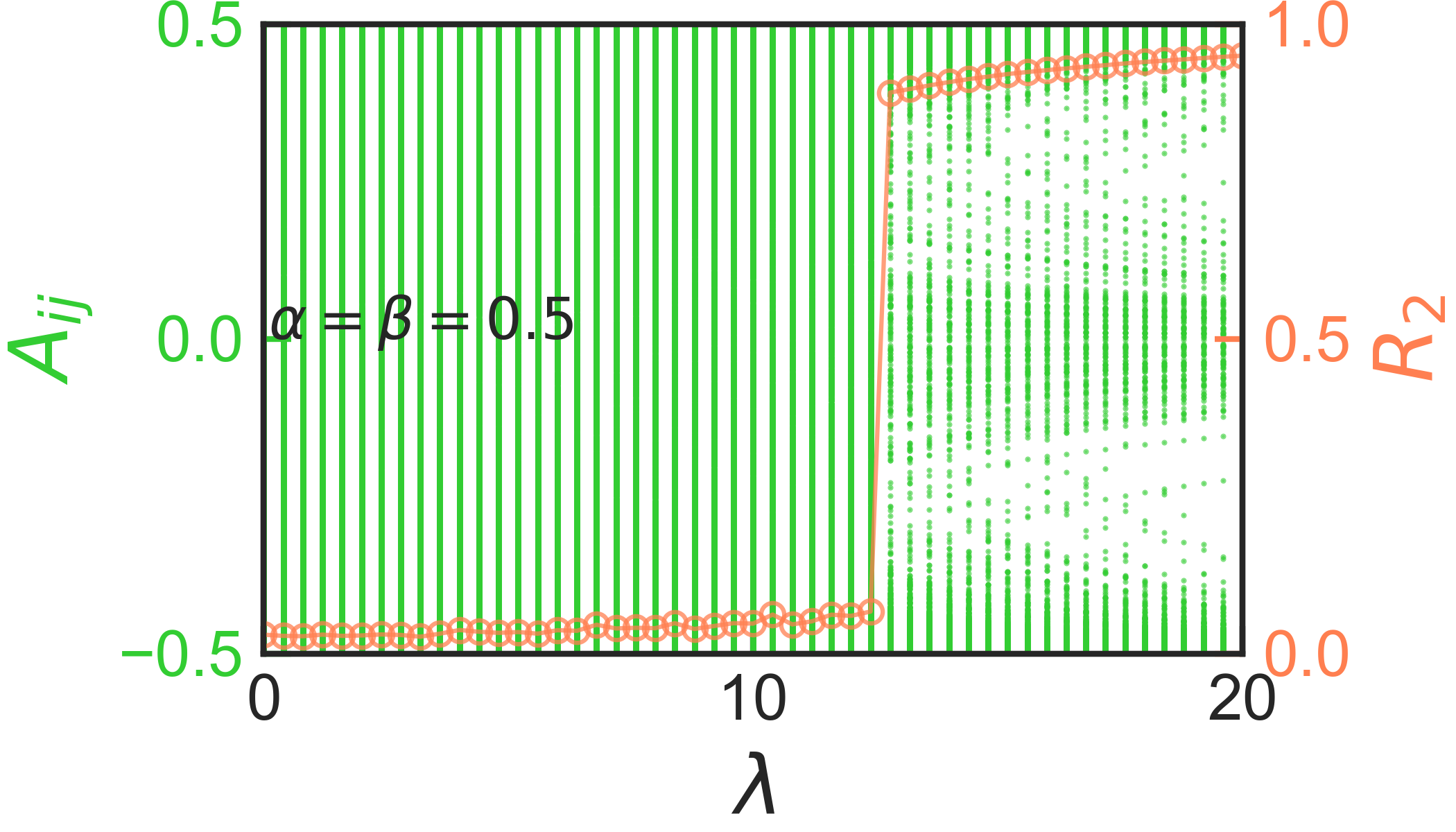}&
	\includegraphics[height=4.5cm,width=6cm]{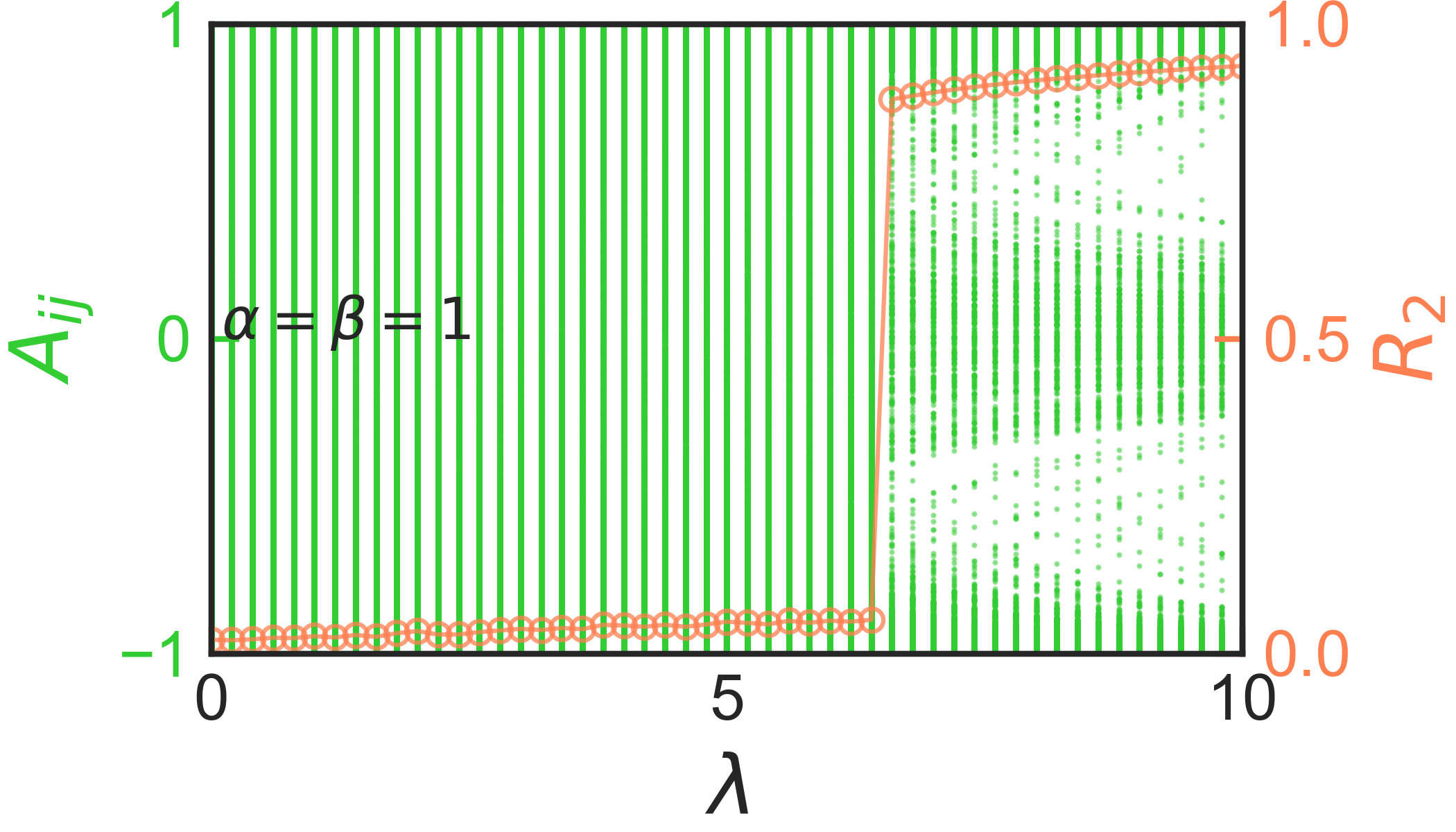}\\
	\includegraphics[height=4.5cm,width=6cm]{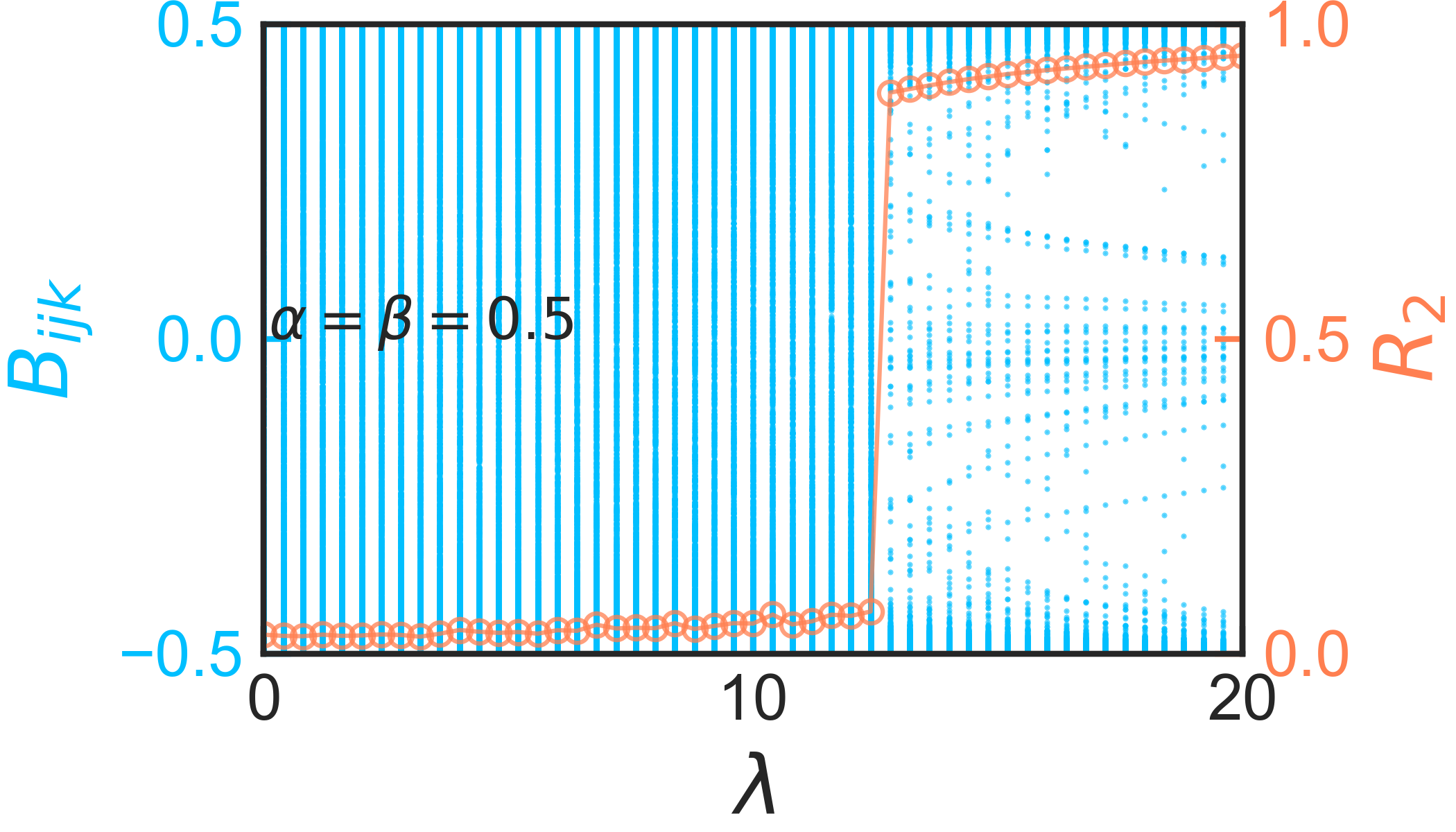}&
	\includegraphics[height=4.5cm,width=6cm]{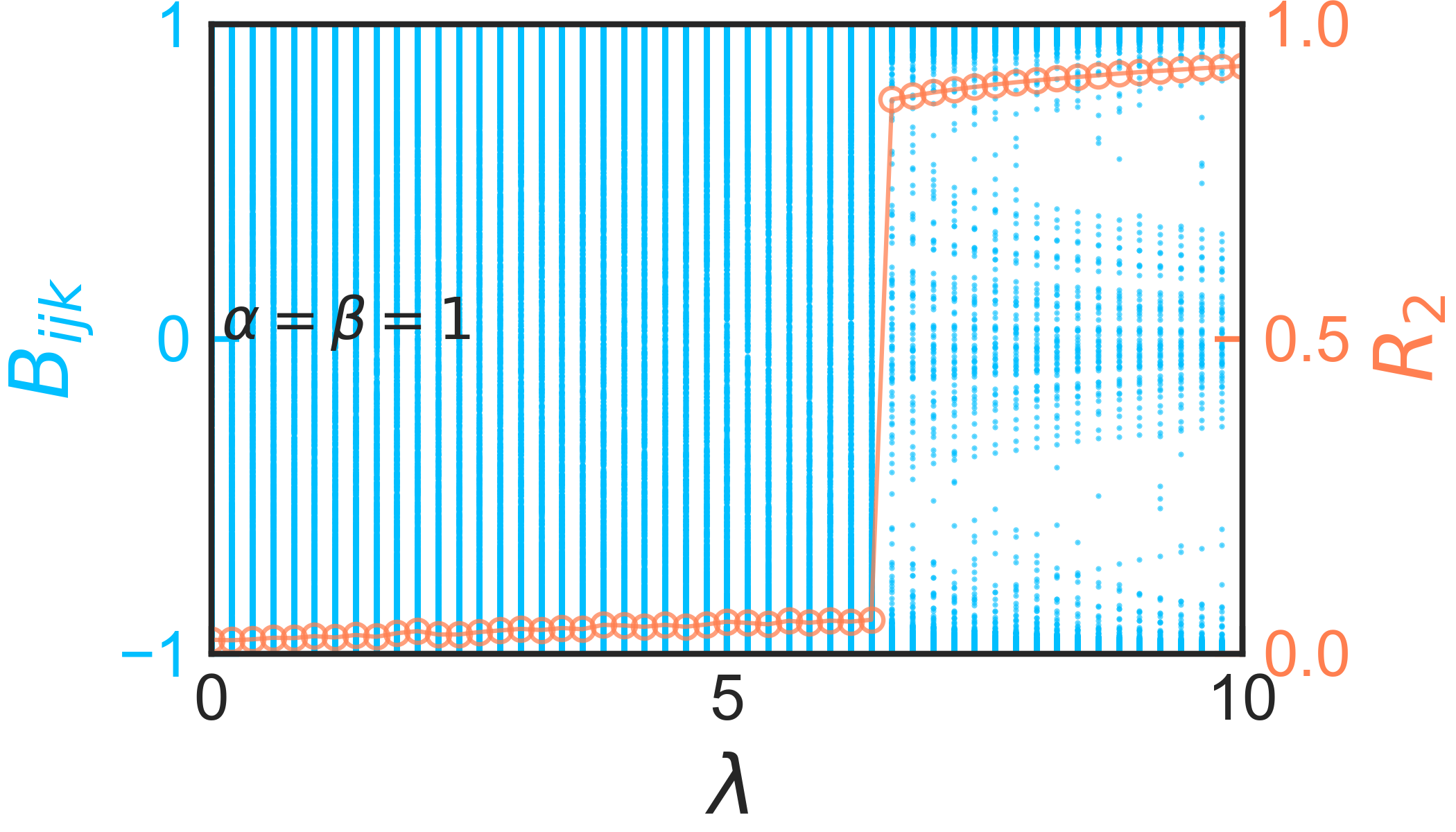}\\
	\end{tabular}{}
	%\vspace{-0.5cm}
	\caption{(Color online) Stationary weights; $A_{ij}$ (dyadic) and $B_{ijk}$ (triadic) plotted against coupling strength $\lambda$ exhibiting transition to synchronization in a random 2-simplicial complex [with mean degrees $\langle k^{[1]}\rangle{=}14$ and $\langle k^{[2]}\rangle{=}10$] simulated for uniform natural frequencies $\omega_i\sim U[-1,1]$, $N=10^3$, $\mu=1$ and $\nu=1$.}
	\label{figureS1}
\end{figure}

 Fig.~\ref{figureS1} depicts the stationary $A_{ij}$ (dyadic) and $B_{ijk}$ (triadic) as a function of coupling strength $\lambda$ exhibiting forward transitions in a random 2-simplicial complex. Both the steady dyadic and triadic weights are bounded within the intervals $[-\alpha/\mu,\alpha/\mu]$ and $[-\beta/\nu,\beta/\nu]$, respectively, in both the incoherent and coherent states.

\subsection{Bistability domains revisited}
\begin{figure}[h!]
	\centering
	\begin{tabular}{cc}
	\includegraphics[height=4.5cm,width=6cm]{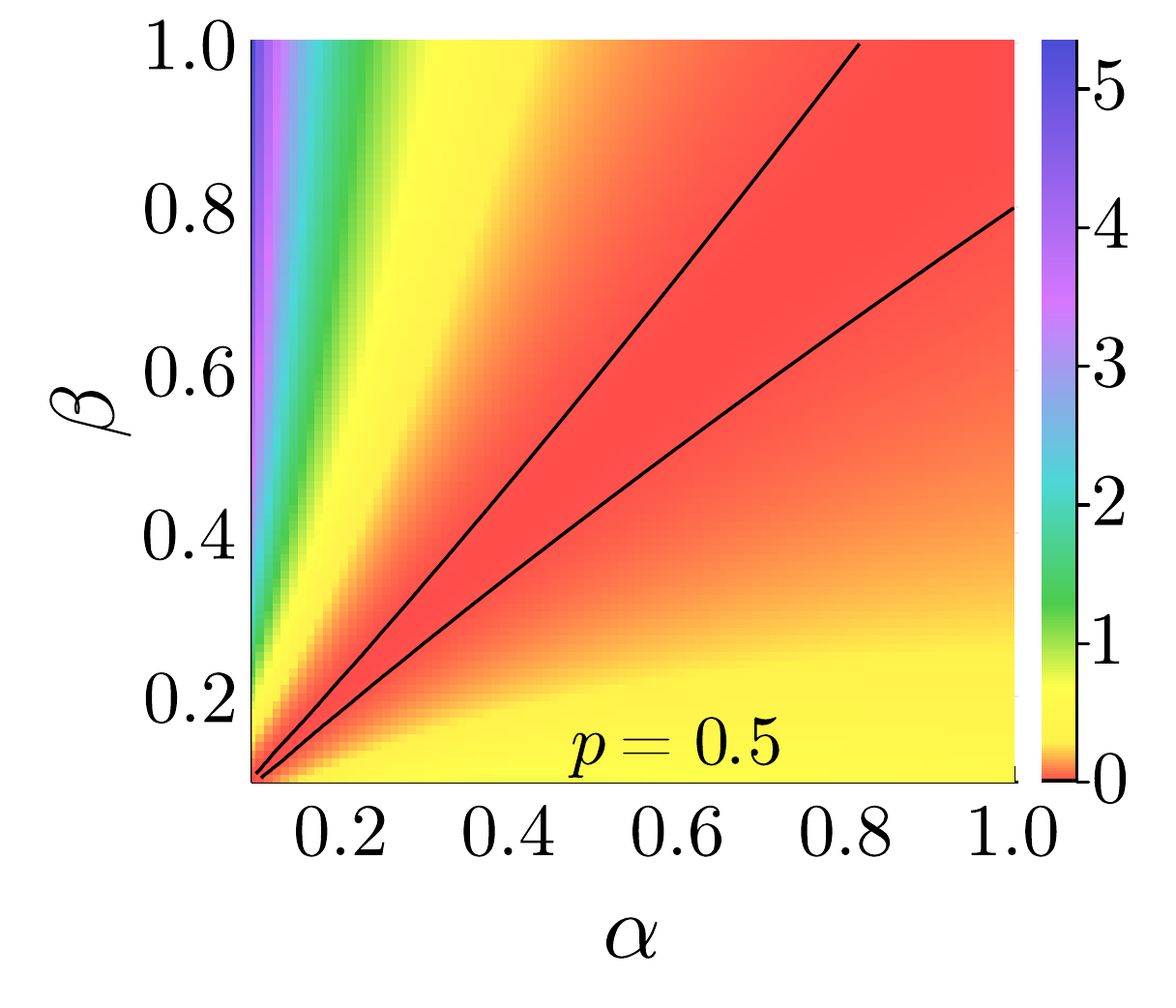}&
	\includegraphics[height=4.5cm,width=6cm]{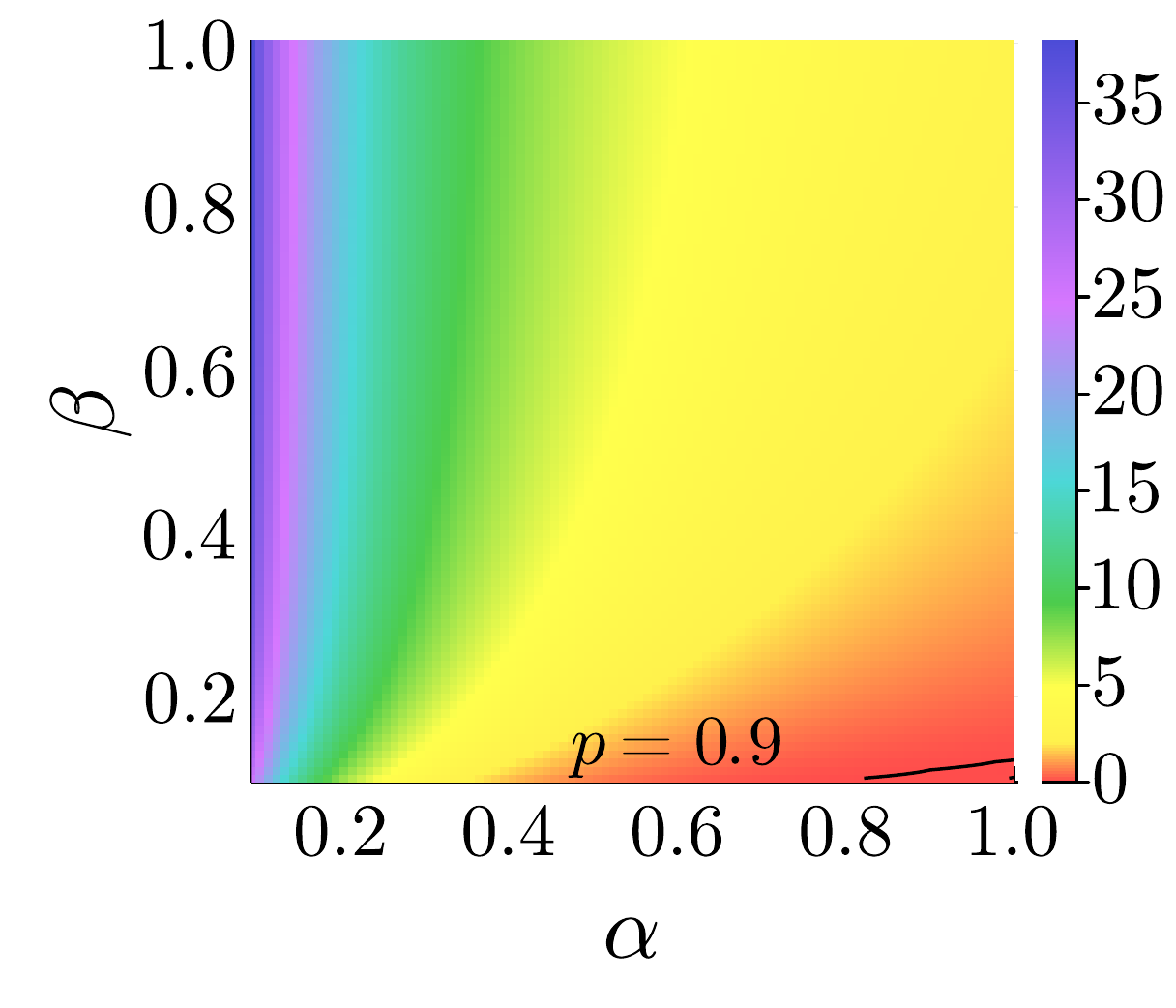}\\
	\includegraphics[height=4.5cm,width=6cm]{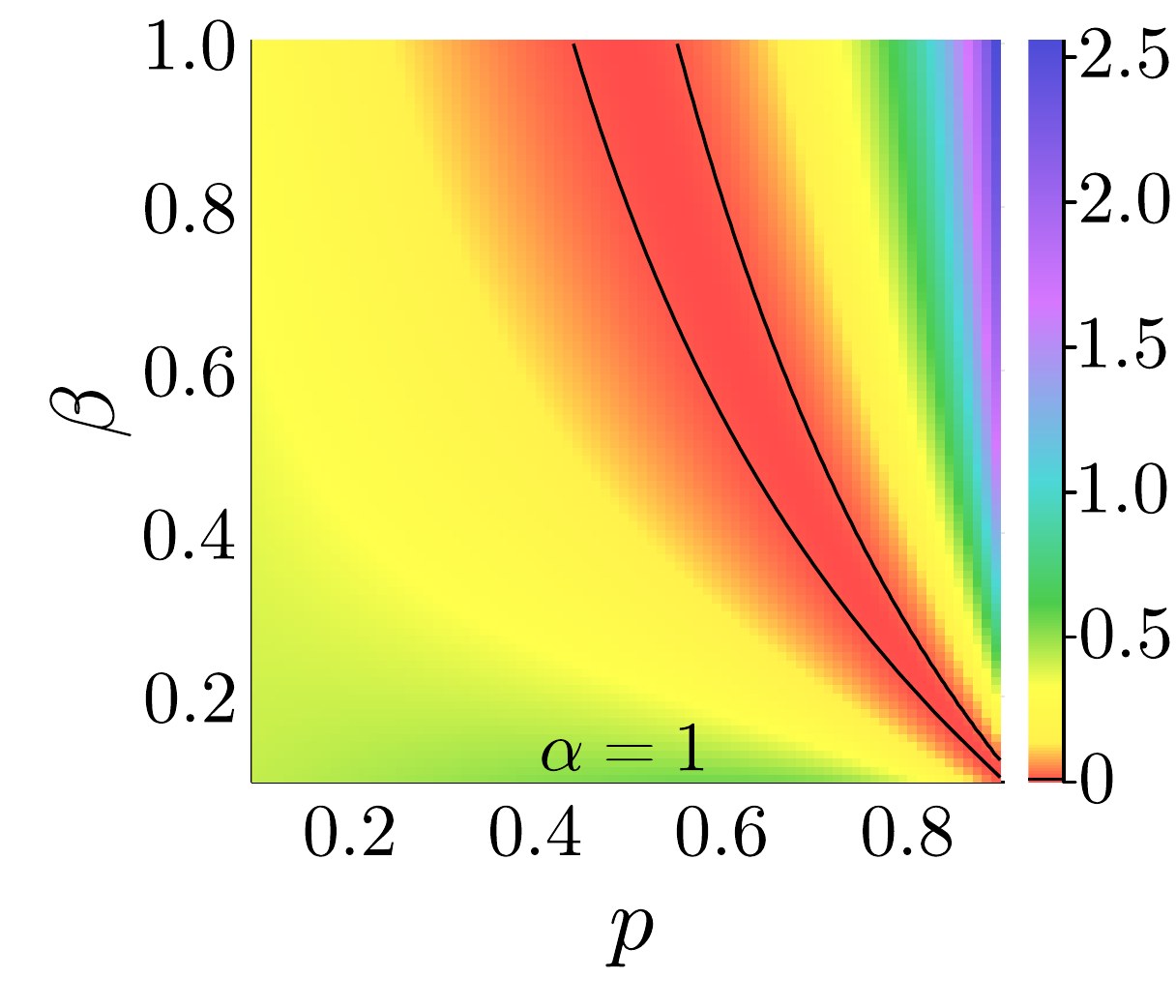}&
	\includegraphics[height=4.5cm,width=6cm]{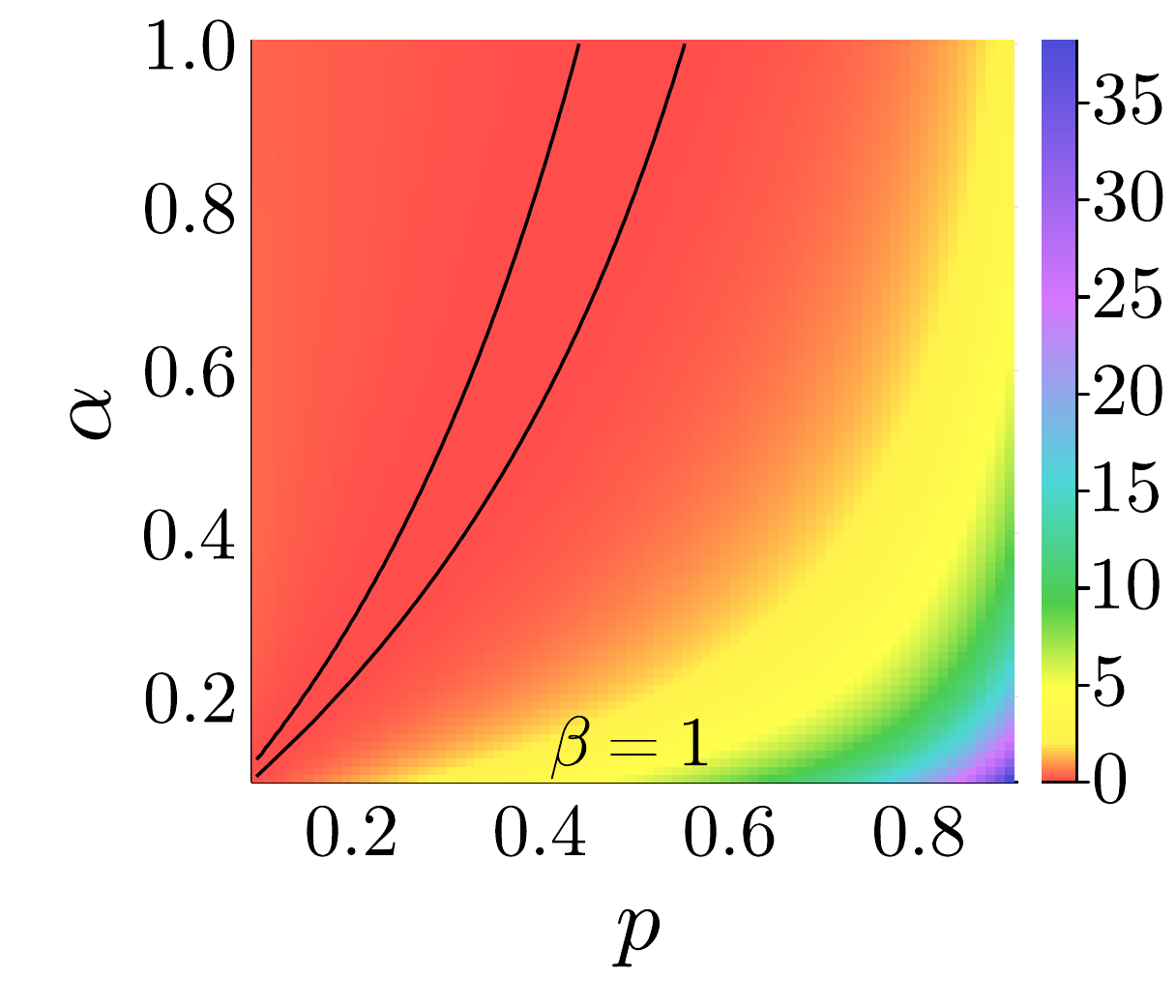}\\
	\end{tabular}{}
	%\vspace{-0.5cm}
	\caption{(Color online) {\em Bistability domains:} Phase diagrams in $(\alpha-\beta)$, $(p-\alpha)$ and $(p-\beta)$ planes representing hysteresis width $|\lambda_f-\lambda_b|$ of transitions in the all-to-all connected 2-simplicial complex simulated for Lorentzian $g(\omega); \Delta=0.1$, $N=10^3$, $\mu=1$ and $\nu=1$.}
	\label{figureS2}
\end{figure}

 In Fig.~\ref{figureS2}, the hysteresis width $|\lambda_f-\lambda_b|$ of transition transpiring in an all-to-all 2-simplicial complex are illustrated in $(\alpha-\beta)$, $(p-\alpha)$ and $(p-\beta)$ planes. The red area between the two black contour lines depicts zeros hysteresis width, representing the occurrence of a second-order transition while the remaining parameter regions manifest an abrupt transition associated with a hysteresis.
